\documentclass[aps,pra,english,twocolumn,showpacs]{revtex4-1}
\def\be{\begin{equation}}
\def\ee{\end{equation}}
\def\bea{\begin{eqnarray}}          
\def\eea{\end{eqnarray}}
\def\bi{\begin{itemize}}
\def\ei{\end{itemize}}
\def\nb{\nonumber}
\usepackage{xcolor}
\usepackage{graphicx}
\usepackage{float}
\usepackage[colorlinks=true, citecolor=blue, urlcolor=blue ]{hyperref}
\usepackage{amsmath,amstext,amssymb,babel,bm,color,times}

\begin{document}

\title{Mimicking quantum correlation of a long-range Hamiltonian by finite-range interactions 
}
\author{ Leela Ganesh Chandra Lakkaraju\(^1\), Srijon Ghosh\(^1\), Debasis Sadhukhan\(^2\), Aditi Sen (De)\(^1\)}
\affiliation{\(^1\) Harish-Chandra Research Institute,  A CI of Homi Bhabha National Institute, Chhatnag Road, Jhunsi, Prayagraj - 211019, India}
\affiliation{\(^2\) School of Informatics, University of Edinburgh,
10 Crichton Street, Edinburgh EH8 9AB, Scotland, United Kingdom}

\begin{abstract}
     The quantum long-range extended Ising model possesses several striking features that cannot be observed in the corresponding short-range model. We report that the pattern obtained from the entanglement between any two arbitrary sites of the long-range model can be mimicked by the model having a finite range of interactions provided the interaction strength is moderate. On the other hand, we illustrate that when the interactions are strong, the entanglement distribution in the long-range model does not match the class of a model with a few interactions. We also show that the monogamy score of entanglement is in good agreement with the behavior of pairwise entanglement. Specifically, it saturates when the entanglement in the finite-range Hamiltonian behaves similarly to the long-range model, while it decays algebraically otherwise.

\end{abstract}

\maketitle

\section{ Introduction } 

Quantum systems with long-range (LR) interactions, naturally emerging in numerous experiments in atomic, molecular, and optical physics \cite{RevModPhys.82.2313,Weber2010,PhysRevLett.108.210401,PhysRevLett.108.215301,Schau2012,Dolde2013,Firstenberg2013,Yan2013,PhysRevLett.107.277201,Britton2012,Islam2011,Islam2013,Richerme2014,Drfler2019,Pagano2019,PhysRevLett.125.133602,Tao2020,RevModPhys.93.025001}, have attracted a great deal of interest in the past decade. Moreover, these systems are often known to possess rich and striking properties which are not typically observed in the models having short-range (SR) interactions. Examples of features include fast
spreading of correlations \cite{Richerme2014,PhysRevB.93.125128,1702.05368,Ares19,PhysRevA.97.062301}, breakdown of the Mermin-Wagner-Hohenberg
theorem \cite{PhysRevLett.17.1133, PhysRev.158.383, PhysRevLett.109.025303}, violation of the area law \cite{PhysRevX.3.031015, PhysRevB.87.035114, RevModPhys.82.277, PhysRevLett.109.267203} and fast state transfer \cite{PhysRevLett.119.170503} to name a few.  
Tremendous advancements in set-ups such as cold atoms in optical lattices, ion traps, and superconducting circuits facilitate quantum control at an unprecedented level, thereby making the simulation of such long-range systems a reality with reasonable system size \cite{RevModPhys.80.885, Bloch2012, Lewenstein2012, RevModPhys.93.025001} and opening up the possibility of practical verification of these interesting characteristics.  

Despite the overwhelming progress in different experimental techniques, the current generation of quantum hardware is not yet scalable. They are far from perfect due to the limited number of controllable qubits and lack of quantum error correction, which are collectively referred to as noisy intermediate-scale quantum (NISQ) hardware~\cite{Preskill18}. Therefore, it is of utmost importance to the current generation of NISQ hardware to use the fewest possible gates so that the noise can be minimal. Quantum variational algorithms like the quantum approximate optimization algorithm (QAOA)~\cite{1411.4028} have been proven to be an efficient tool to simulate many-body systems~\cite{Pagano2019,Medvidovi2021,Tao2020,2109.10909,1801.03922,PhysRevX.9.031006} which are also suitable for NISQ hardware~\cite{Cerezo2021,matos2022,referee1,referee2,referee3}. However, to simulate an end-to-end connected LR system (cf. \cite{qaoa_lri}, in which LR systems are also simulated via the QAOA), if we use only a single two-qubit gate per interaction, we require at least ${O}(N!)$ two-qubit gates for an $N$-site system, which again needs to be optimized over multiple iterations. In gate-based quantum hardware, the two-qubit gates in general introduce more noise in the system than the single-qubit ones and have an overall low fidelity~\cite{Boixo2018}. Therefore, an exponential use of two-qubit gates can make the overall simulation too noisy to obtain any meaningful result. In this work, we try to circumvent this problem by approximating a LR Hamiltonian with a finite number of pairing interactions, keeping the overall behavior of two-qubit entanglement behavior intact, which in turn results in an exponential-to-polynomial reduction of the usage of two-qubit gates.


In an LR model, the two-body interaction potential decays algebraically with their relative distance, typically like ${1}/{r^{\alpha}}$, where $r$ is the relative distance between the two bodies and the exponent $\alpha$ controls the strength of the interaction. For such a system of spatial dimension $D$, the interactions are strong when $0 \le \alpha \le D$ while those with $\alpha > D+1$ are called weak. The weak LR interactions effectively behave like the SR ones where the correlations have an exponential tail except at the critical point, where the correlations are algebraic. 
On the other hand, when $\alpha < D+1$, 
the correlations always have an algebraic tail regardless of the critical points. This is clearly a very distinctive feature of a true LR Hamiltonian, having counter-intuitive characteristics. From the point of view of quantum correlations, although information-theoretic measures \cite{RevModPhys.84.1655,Bera2017} may show long-range order~\cite{Ciliberti2010,Tomasello2011,Maziero2012,Sadhukhan2016}, measures from the entanglement-separability paradigm are typically short range when $\alpha>D+1$. In this work, we concentrate on the regime when $\alpha<D+1$ where classical correlations always have an algebraic tail, irrespective of the critical point.

Typically, in an SR model, entanglement follows the area law when the ground state is gapped and has a finite range of interactions \cite{RevModPhys.82.277}. Indeed, entanglement entropy in the ground state of a one-dimensional gapped system saturates to a constant value in the thermodynamic limit as an implication of the area law~\cite{0705.2024}, while the same grows logarithmically if the system is a gapless one~\cite{hep-th/0405152, quant-ph/0503219}. On the other hand, in the presence of long-range interactions, these results are not valid anymore and entanglement can grow logarithmically even away from the critical point~\cite{1506.06665, PhysRevA.97.062301}. In fact, under certain special circumstances, LR interactions allow sublogarithmic growth of entanglement entropy \cite{1405.2804,1611.08506,1804.06357,Ares19} or even as a volume law \cite{1401.5922} which is a clear violation of the area law. 
Long-range systems, where area law is not valid, should in principle not be efficiently simulable with numerical tools like tensor networks~\cite{1306.2164}. However, it has been shown that existing numerical tools such as matrix product states~\cite{PhysRevLett.109.267203,VUMPS,1801.00769,Laurent2018, Laurent2021, Laurent2021, PhysRevX.3.031015,prx1, prb1, prb2, pra1} can produce a good match with the exact results. This may be attributed to the analysis of the distribution of entanglement
which can be mimicked with long but not infinite-range interactions. We show that this is indeed true in the intermediate regime $1<\alpha<2$ as far as the two-qubit entanglement is concerned.      

Besides entanglement entropy, the two-site entanglement between arbitrary pairs of spins is directly related to applications such as secure quantum communication and quantum internet, involving multiple parties.
Since many-body quantum systems are often considered to be promising premises to generate multipartite entangled states, a long-range Hamiltonian can be more useful than the SR ones. In this article, we, therefore, investigate the two-qubit entanglement between different lattice sites over the entire spin chain. 
Note that, unlike classical correlations, an algebraic decay of two-site entanglement with distance is restricted by the monogamy of entanglement \cite{coffman2000,ckw2,monorev}. 
In this work, we address the following questions:

\emph{ Can a ground state of an end-to-end fully connected LR Hamiltonian be efficiently simulated by a finite number of pairing interactions?} 

\emph{If so, how many neighbors are required to mimic the same behavior of entanglement in the ground state and how does that number vary with the exponent $\alpha$ i.e., the strength of the interaction?}

These questions are especially relevant when we wish to create a link between the different hardware platforms that are presently available. For example, in an ion-trap simulator, simulation of SR spin models is challenging, while the models realized are typically long range with a high enough exponent, thereby possessing vanishing long-range behaviors \cite{Islam2011,Islam2013}. On the other hand, in a gate-based simulator with superconducting qubits, e.g., in IBM, Google, and Rigetti, the simulation of an LR model is problematic since the simulation becomes extremely noisy due to the exponential use of two-qubit gates. 
Therefore, it would be tremendously helpful if we could simulate the entanglement content of an end-to-end LR model with a system having  fewer neighbor interactions 
so that the corresponding
Hamiltonian can act as a representative between the two simulators. 

In this paper, by choosing a family of LR models in one dimension that can be solved analytically, we show that at the quasi-local regime having moderate interaction strength, we can reproduce nearly the same pattern for two-qubit entanglement of the ground state with a model possessing a few finite-neighbor interactions. In particular, when $1<\alpha<2$, where classical correlations are known to have an algebraic decay, pairwise entanglement is mostly short-range and can be mimicked by a finite-neighbor Hamiltonian. We also illustrate that the number of neighboring interactions can be further reduced if we allow stronger interaction strength in the few-neighbor Hamiltonian compared to the target Hamiltonian with an end-to-end connection.  However, in the non-local regime $\alpha<1$, we observe that entanglement can also have an algebraic tail $E_r \approx 1/r^{\alpha}$ and therefore one requires a pairing interaction of the order of the size of the system (approximately equal to $N$) to reproduce nearly the same entanglement pattern as the true LR model. We supplement our results by analyzing the monogamy of entanglement in the ground state and argue that in the regime where the monogamy score tends to saturation with the increase in the range of interactions, a finite-neighbor interaction can be a good representative of the true LR model for mimicking the trends of pairwise entanglement.


The paper is organized as follows. In Sec.\ \ref{sec:model} we introduce a family of Hamiltonian that we deal with and include a summary of the diagonalization procedure to make the paper self-contained. The critical points are discussed in Sec.\ \ref{sec:critical}. 
Section\ \ref{sec:correlation} includes a short summary of the evaluation process for correlations that are required to compute entanglement. In Sec. \ref{sec:gs_ent} we manifest scenarios where a finite range of interactions is enough to produce the pattern of two-qubit entanglement in the fully connected LR models. Section  \ref{sec:ent_phase} reports the change in the behavior of entanglement depending on the phases. In Sec.\ \ref{sec:monogamy} we investigate the monogamy score of entanglement in these systems and argue that the trends of monogamy can also indicate whether a few interactions can mimic entanglement patterns of the LR models. We summarize in Sec.\ \ref{sec:conclu}.  

 \section{The family of long-range models}
 \label{sec:model}

We introduce the model Hamiltonian under consideration and briefly describe the diagonalization procedure. 

 

We consider an Ising-type model with long-range interacting terms. Variations of these models have already been studied in recent literature~\cite{Vodola1, Vodola2, Maity_2019} and were shown to have contrasting properties as compared to the SR models. 
The LR Hamiltonian of $N$ sites reads
\bea
H=
\sum_{n=1}^N
\left[
 \frac{h'}{ 2}\sigma^z_n +
\sum_{r=1}^{\cal Z} 
J_r^{\prime}\sigma^x_n\prod_{i=n+1}^{n+r-1}\sigma^z_{i}\sigma^x_{n+r}
\right],
\label{Hamil_fLR}
\eea 
with open boundary conditions, where \(\sigma^\alpha\) (\(\alpha =x, y, x\)) are the Pauli matrices. Here $h$ is the transverse magnetic field and $J_r^{\prime}=\frac{J}{A} \frac{1}{r^{\alpha}}$ is the interaction strength depending on the distance $r$ between the sites, with the exponent $\alpha$ the tuning parameter, which dictates the interaction strengths between different spins, and $A$ a constant. We set \(h'/J = h\) and $J_r'/J = J_r$. When $\alpha \sim 0$, the model behaves like the LR Ising model, which is similar to the Lipkin-Meshkov-Glick (LMG) model ~\cite{LMG, Morigi_2018} while for $\alpha>2$, the model increasingly resembles the nearest-neighbor Ising model~\cite{Luca2013, Eisert2013,Cevolani2016} with increasing $\alpha$ values and therefore falls within the universality class of the quantum transverse Ising model. The value of ${\cal Z}$ sets the number of pairwise interactions per site in the lattice, which is also called the coordination number. For example, with ${\cal Z} = 1$, we get the nearest-neighbor (NN) Ising model, while ${\cal Z} = 2$ represents the next-nearest-neighbor extended Ising model, and the true LR extended Ising model occurs with ${\cal Z} = N-1$. Any intermediate values of ${\cal Z}$ correspond to the few-neighbor extended Ising models. We expect to reveal contrasting entanglement patterns in two distinct regions: (i) $\alpha < 1$, which we call the non-local regime, and (ii) $1< \alpha < 2$, referred to as the quasilocal regime. We compare the results with the local regime ($\alpha > 2$) results which belong to the SR Ising universality class.

In this model, we notice that pairwise interaction terms between $i$ and $j$ have the form $\sigma^x_i\sigma^z_{i+1}\ldots\sigma^z_{j-1}\sigma^x_{j}$ instead of $\sigma^x_i\sigma^x_{j}$, which allows the Hamiltonian to be treated analytically. However, within the truncated Jordan-Wigner (JW) approximation~\cite{Jaschke_2017}, both models are the same. The constant in $J_{r}$ which can be considered as normalization is $A = \sum_{r=1}^{\mathcal{Z}} r^{-\alpha}$, which fixes the ferromagnetic critical point at $h=2$. For a finite LR system of size $N$, $A$ evaluates to $H_{N-1}^{(\alpha)}$, the generalized harmonic number, which in the thermodynamic limit becomes the Riemann zeta function $\zeta(\alpha)$. For any of the few-neighbor extended Ising models $A =H_{{\cal Z}}^{(\alpha)}$. 

\subsection{Few-neighbor extended Ising model}


We investigate the behavior of entanglement in models with several few-neighbor pairwise interactions instead of studying the entanglement properties of the LR model. Specifically, except $\mathcal{Z} = 1$ (NN Ising) or $\mathcal{Z} = N-1$ (true LR), we study all other $\mathcal{Z}$ values, thereby dealing with $\mathcal{Z}$-neighbor extended Ising models. As we will show, after a certain $\mathcal{Z}$ value, quantum correlations can mimic the behavior obtained for the LR model.

\subsection{Long-range extended Ising model}

For any finite-size system, the end-to-end connection is considered when $\mathcal{Z} = N-1$, and therefore we call the same as the true long-range extended Ising model or simply the LR extended Ising model. 
In the thermodynamic limit, i.e., $N\to\infty$ the  LR Hamiltonian can only be normalized when $\alpha>1$ so that $\sum_r J_r =1$. In this case, the Hamiltonian takes the form 
\bea
H=
\sum_{n=1}^{\infty}
\left[
 \frac{h}{2} \sigma^z_n +
\frac{1}{\zeta(\alpha)}\sum_{r=1}^{ {\infty}} 
\frac{1}{r^{\alpha}}\sigma^x_n\prod_{i=n+1}^{n+r-1}\sigma^z_{i}\sigma^x_{n+r}
\right],
\label{Hamil_LR}
\eea
where the normalization is given by $A = \zeta(\alpha)$, the Riemann zeta function. 
For $0 <\alpha \le 1$, the normalization does not exist in the thermodynamic limit, and hence, we must restrict ourselves to a finite-size system in order to maintain the normalization. Derivatives of this Hamiltonian have already been studied \cite{Vodola1, Vodola2, Jaschke_2017, AmitDutta2017, Cevolani2016, Maity_2019, SonicHorizon, LengthInLR} in the literature. Note that the presence of $\sigma^z$-string operators in the pairwise interaction term makes the model different from the LR Ising model. Within the truncated Jordan-Wigner approach~\cite{Jaschke_2017}, where we truncate the fermionic operator up to quadratic order, the LR Ising model reduces to the LR extended Ising model and can be treated analytically. In general, the truncation approximation becomes better deep in the disordered phase where they satisfy $\langle \sigma^z_i \rangle = 1-2c_i^\dagger c_i \approx 1$.

\subsection{Diagonalization}

Let us now illustrate the procedure by which a $\mathcal{Z}$-neighbor extended Ising model can be diagonalized analytically. Due to the specific nature of the pairwise interaction in the long-range interaction terms of the Hamiltonian, these families of Hamiltonian can be mapped to quadratic free-fermion models which can be solved analytically. Here we limit ourselves to the $+1$-parity subspace of the Hilbert space \cite{lieb1961, Katsure1962, glen2020}; note that $H$ commutes with the parity operator $P = \prod_{n=1}^N\sigma_n^z$. The first step in the diagonalization is to apply  the Jordan-Wigner transformation, given by

\bea
&&
\sigma^x_n~=~
 -
 \left( c_n + c_n^\dagger \right)
 \prod_{m<n}(1-2 c^\dagger_m c_m)~,\\
 \label{sigma_x}
&&
\sigma^y_n~=~
 i
 \left( c_n - c_n^\dagger \right)
 \prod_{m<n}(1-2 c^\dagger_m c_m)~, \\
 \label{sigma_y}
&&
\sigma^z_n~=~1~-~2 c^\dagger_n  c_n~, 
\label{JordanWigner}
\eea
where fermionic operators $c_n$ satisfy
$\left\{c_m,c_n^\dagger\right\}=\delta_{mn}$ and 
$\left\{ c_m, c_n \right\}=\left\{c_m^\dagger,c_n^\dagger \right\}=0$.
For periodic boundary condition, the Hamiltonian $H$  becomes \cite{lieb1961}
\be
 H~=~P^+~H^+~P^+~+~P^-~H^-~P^-~,
\label{Hc}
\ee
where
$
P^{\pm}=\frac12\left[1\pm P\right]
$
are projectors on subspaces with even ($+$) and odd ($-$) parity,
\be 
P~=~
\prod_{n=1}^N\sigma^z_n ~=~
\prod_{n=1}^N\left(1-2c_n^\dagger c_n\right),
\label{P}
\ee
and  
$
H^{\pm}
$
are the corresponding reduced Hamiltonian. Although the spin Hamiltonian is periodic, after the JW transformation, the $c_n$ in $H^-$ satisfy
periodic boundary conditions, i.e., $c_{N+1}=c_1$, while the $c_n$ in $H^+$
are anti periodic, i.e., $c_{N+1}=-c_1$.

When dealing with the periodic Hamiltonian in the thermodynamic limit, we constrain ourselves to the positive-parity subspace, and the Hamiltonian in Eq. (\ref{Hamil_LR}) reads 
\bea
H^+ &=& \sum_n \frac h2\left(1-2c_n^\dag c_n\right) \nb \\
&&+ \sum_{n,r} J_r\left[ (c_n^\dag c_{n+r}-c_nc_{n+r}^\dag) + (c_n^\dag c_{n+r}^\dag-c_nc_{n+r})\right] \nb, \\
\label{HcLR}
\eea
with the anti-periodic boundary condition $c_{N+r}=-c_r$ $\forall ~ r$, that corresponds to the case where the total number of quasi particles is even, i.e., $\sum_{n=1}^N c_n^\dag c_n = s $ is even so that $\prod_{n=1}^N (1-2c_n^\dag c_n)=(-1)^s$. 
%

In the thermodynamic limit, the translationally invariant $H^+$ is diagonalized by a Fourier transform followed by a Bogoliubov transformation \cite{bm1, bm2, glen2020}. 
The Fourier transform applicable for the anti-periodic boundary condition  is given by  
\be
c_n~=~ 
\frac{e^{-i\pi/4}}{\sqrt{N}}
\sum_k c_k e^{ikn}~,
\label{Fourier}
\ee
where the pseudomomentum takes half-integer values as follows,
\be
k = \frac{(2m - 1)\pi}{N}, \text{where } m = 1, 2, \ldots (N-1).
\label{halfinteger}
\ee

In the case of even parity, $P = \prod_{n=1}^N\sigma_n^z$, the fermionic creation (annihilation) operators satisfy anti-periodic boundary conditions. Using the Fourier transformation given in Eq. (\ref{Fourier}), we can rewrite the Hamiltonian as
\bea
H^+ &=&
2\sum_{k>0} 
\left(\frac{h}{2} -  \mbox{Re}(\tilde{J_k})\right) \left(c_k^\dag c_k+c_{-k}^\dag c_{-k}\right) \nb \\
&&+  ~\mbox{Im}(\tilde{J_k})\left(c_{k}^\dag c_{-k}^\dag + c_{-k}c_{k}\right) -\frac{h}{2},
\label{HkLR}
\eea
where $\tilde{J_k}=\sum_{r=1}^\mathcal{Z} J_r e^{ikr}$ is the Fourier transform of $J_r$. 
Therefore, we have $\tilde{J_k}=\frac{1}{H_{{\cal Z}}^{(\alpha)}}\sum_{n=1}^{\cal Z}\frac{x^n}{n^\alpha}$.

\begin{figure}[h]
    \centering
    \includegraphics[width=9cm,height=13.5cm]{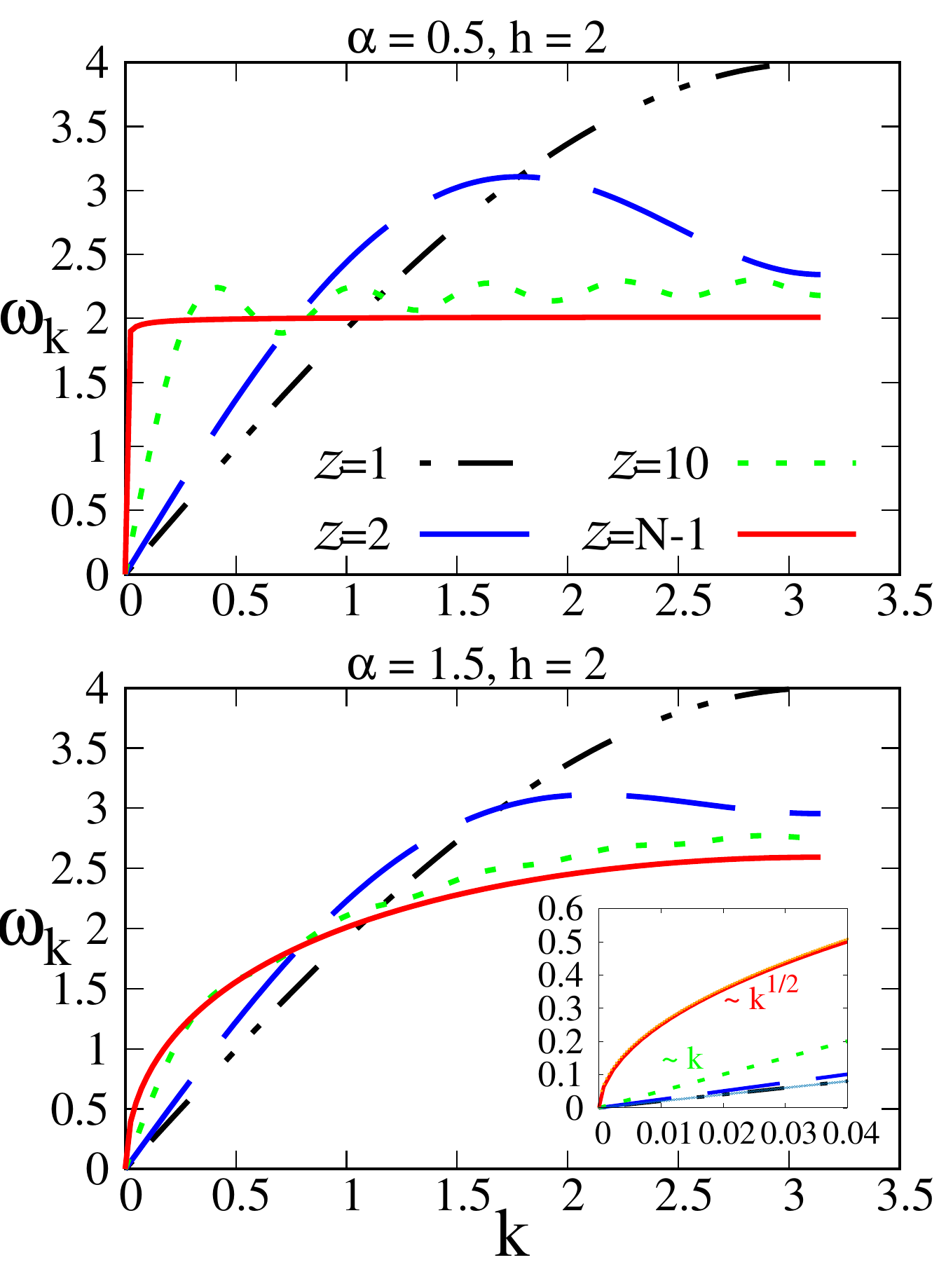}
    \caption{(Color Online.) \textbf{Dispersion relation to prove criticality.} $\omega_{k}$ (ordinate) vs. $k$ (abscissa) at the critical point, $h_{c}^{1} = 2$. Different lines indicate different values of $\mathcal{Z}$. Lower and upper panels depict quasi-local and non-local regimes respectively. Specific choices of parameters are mentioned in the headings of each plot. The inset indicates the dependence of $k$ on the velocity of the quasi-particles for the LR model and a few-neighbor model. Here $N = 256$ and therefore the line plots represent the joining lines of $N$ discrete momenta (throughout the paper, we consider $N = 256$ unless stated otherwise). Both axes are dimensionless.} 
    \label{fig:dispersion}
\end{figure}

The stationary Bogoliubov-de Gennes equations are
\bea
\omega_k
\left(
\begin{array}{c}
U_k \\
V_k
\end{array}
\right)=
2\left[\sigma^z(\frac{h}{2}-\mbox{Re}(\tilde{J_k}))
 +\sigma^x\mbox{Im}(\tilde{J_k})\right]
\left(
\begin{array}{c}
U_k \\
V_k
\end{array}
\right),~~~~,
\label{BdGkLR}
\eea
with eigenfrequencies
\be
\omega_k=2\sqrt{(\frac{h}{2}-\mbox{Re}(\tilde{J_k}))^2+ \mbox{Im}(\tilde{J_k})^2},
\label{omegakLR}
\ee where $\mbox{Re}(\tilde{J_k})=\sum_{r=1}^\mathcal{Z} J_r \cos{kr} $ and $ \mbox{Im}(\tilde{J_k}) = \sum_{r=1}^\mathcal{Z} J_r \sin{kr}$ (see Fig. \ref{fig:dispersion}).
%
%
%
%
%
Here $(U_k,V_k)$ and $(-V_k,U_k)$ are the corresponding eigenvectors. We can now define a new quasiparticle 
\be
\gamma_k = U_k c_k + V_{-k}c_{-k},
\ee
 which finally brings the Hamiltonian to its diagonal form, 
 \be
 H^+ = E_0 + \sum_k \omega_k\gamma_k^\dagger\gamma_k, 
 \ee 
where $E_0 = -\frac{1}{2}\sum_{i=1}^N \omega_i$.

Let us briefly discuss here the diagonalization procedure of the finite size Hamiltonian of both few-neighbor and true LR Hamiltonian with open boundary conditions. 
We first rewrite the Hamiltonian in Eq. (\ref{HcLR}) as
\begin{equation}
H= \sum_{r = 1}^{N-1} \sum_{i, j=1}^{N} \left(c_{i}^{\dagger} A_{i j} c_{j}+ c_{i}^{\dagger} B_{i j} c_{j}^{\dagger}+\text { H.c. }\right), 
\label{A_B_hamiltonian}
\end{equation}
where $A_{ij}$ and $B_{ij}$ are the $ij$ th elements of symmetric and anti-symmetric matrices, respectively, having dimension $N \times N$, given by,
\begin{equation}
\begin{aligned}
A_{i j} &=-h \delta_{i j}+ J_{r} \delta_{i+r, j}+ J_{r} \delta_{i, j+r}, \\
B_{i j} &= \left(J_{r} \delta_{i+r, j} - J_{r} \delta_{i , j+r}\right).
\label{A_B}
\end{aligned}
\end{equation}
 Here, for the finite case, we consider open boundary conditions because otherwise, the effective maximum distance between two lattice sites becomes $N/2$ instead of $N$. We diagonalize the Hamiltonian in Eq. (\ref{A_B_hamiltonian}) with linear transformations, which take care of both the Fourier and Bogoliubov transformations at the same step and are given by 
\begin{equation}
\begin{aligned}
\mu_{k} &=\sum_{i=0}^{N-1}\left(d_{k i} c_{i}+e_{k i} c_{i}^{\dagger}\right), \\
\mu_{k}^{\dagger} &=\sum_{i=0}^{N-1}\left(d_{k i} c_{i}^\dagger+e_{k i} c_{i}\right),
\label{bogo}
\end{aligned}
\end{equation}
where $k = 0, 1, 2, \ldots, N-1$ and $d_{ki}, e_{ki} \in \mathbb{R}$ to be found numerically such that the Hamiltonian becomes diagonal $H=\sum_k \xi_k \mu_{k}^{\dagger}\mu_{k}+ \mbox{const}$.  
Since $\mu_{k}$ obeys the fermionic anti-commutation relations $\{\mu_{k},\mu_{k^{'}}\} = \delta_{k,k^{'}}$, we can also write Eq. (\ref{HcLR}) in terms of $\mu_{k}$ such that the coupled equations
\begin{equation}
\begin{aligned}
&(A+B) \phi_{k}^{T}=\xi_{k} \psi_{k}^{T},\\
&(A-B) \psi_{k}^{T}=\xi_{k} \phi_{k}^{T}
\label{A+B}
\end{aligned}
\end{equation}
hold. 
The coefficients are then found by solving the linear matrix equations,
\begin{equation}
\begin{aligned}
(A+B)(A-B) \psi_{k}^{T}=\xi_{k}^{2} \psi_{k}^{T},\\
(A-B)(A+B) \phi_{k}^{T}=\xi_{k}^{2} \phi_{k}^{T},
\label{(A+B)(A-B)}
\end{aligned}
\end{equation}
where $\psi_{k}$ and $\phi_{k}$ are 
$\phi_{k i} =d_{k i}+e_{k i} $ and
$\psi_{k i} =d_{k i}-e_{k i}$.
%
When $\xi_{k} \neq 0$, we first evaluate $\phi_{k}^{T}$ from Eq. (\ref{(A+B)(A-B)}) and then $\psi_{k}^{T}$ is obtained from Eq. (\ref{A+B}), while for $\xi_{k} = 0$, it is possible to compute both $\psi_{k}^{T}$ and $\phi_{k}^{T}$ by solving Eq. (\ref{(A+B)(A-B)}) where their relative sign remains arbitrary.  



\begin{figure}[h]
    \centering
    \includegraphics[width=9cm,height=12.5cm]{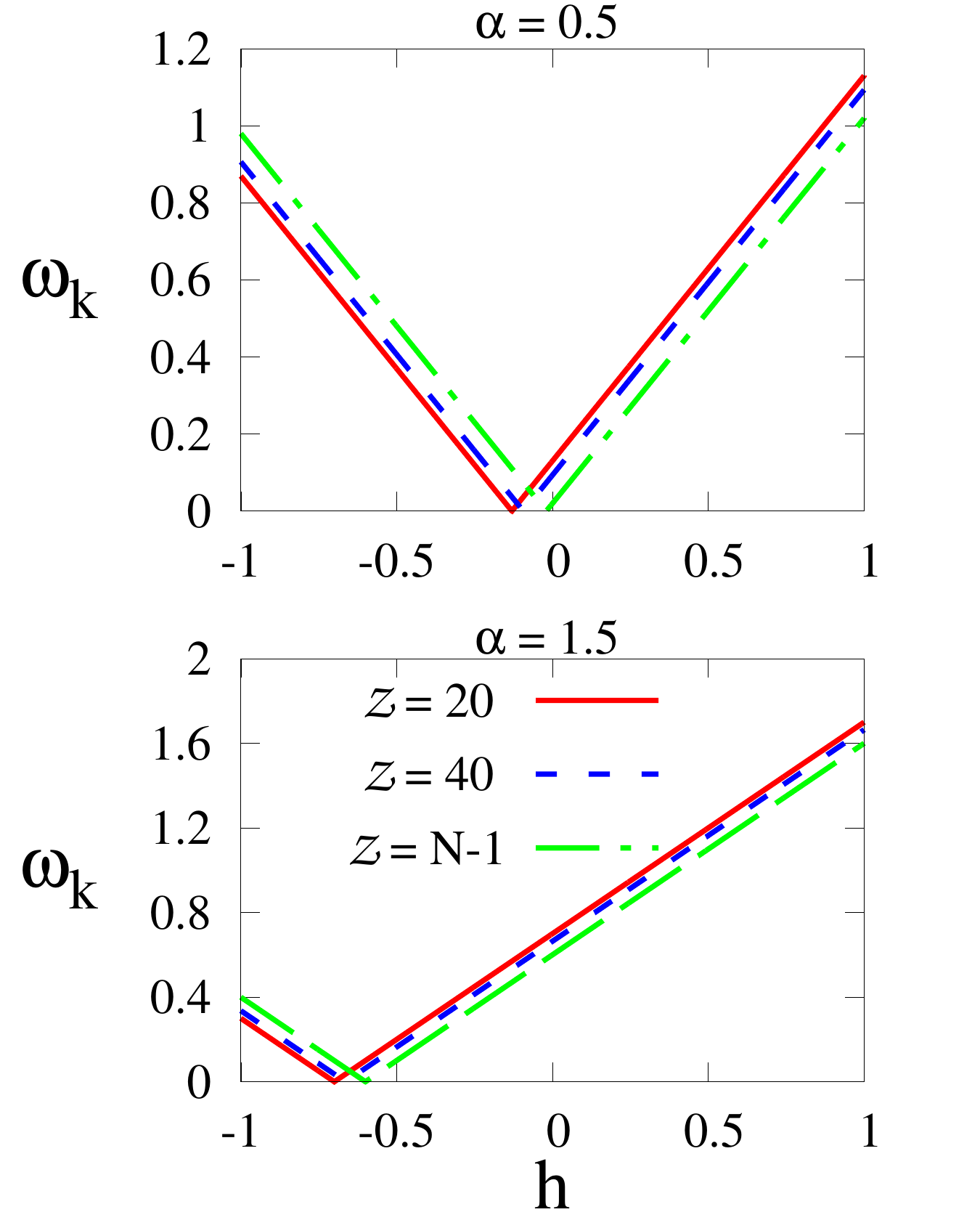}
    \caption{(Color Online.) {\bf{Dispersion relation to find critical point.}} eigenfrequency $\omega_{k}$ (vertical axis) vs $h$ (horizontal axis) at $k = \pi$. The value of $h$ for which $\omega_k$ becomes zero is the critical point, $h_c^2$. It is in good agreement with Eq. (\ref{Eq: hc2}). Different lines correspond to different values of $\mathcal{Z}$. The upper and lower panels are for non-local and quasi-local regimes respectively. Both axes are dimensionless.}
    \label{fig:gap_closing}
\end{figure}

\section{Critical points of a few-neighbor extended Ising model}
\label{sec:critical}
We now determine the quantum phase transitions for the family of LR Hamiltonian \cite{Vodola1,Vodola2,Maity_2019,1506.06665}. Such an analysis also helps us identify critical points and obtain the dispersion relation for the few-neighbor extended Ising model.  The critical point of the true LR model in the thermodynamic limit can be found easily from Eq.~(\ref{omegakLR}) where the gap closes $(\omega_k = 0)$ at $h_c^1=2$ and $k=0 ~ \forall~ \alpha>1$. However, for $\alpha <1 $, $h_c^1=2$ does not remain a critical point in the thermodynamic limit (since the normalization $A=\sum_r J_r $ fails). However, for finite-size systems, we can always have the normalization, and therefore $h_c^1=2$ continues to be a quantum critical point. The other critical point corresponding to $k=\pi$ is located at
\begin{equation}
    h_c^2=\frac{2}{\zeta(\alpha)}\sum_r \frac{(-1)^{r}}{r^\alpha}=2({2^{1-\alpha}-1}),
\end{equation}
which continues to exist even at the thermodynamic limit irrespective of the values of $\alpha$.

Let us now concentrate on the critical point of the $\mathcal{Z}$-neighbor extended Ising model. By analyzing the vanishing $\omega_k$, i.e., the points where the gap vanishes, the critical point corresponding to $k=0$ is again at $h_c^1 = 2$, while the critical point with $k=\pi$ gets shifted to   
\begin{equation}
    h_c^2=\frac{2}{H_{{\cal Z}}^{(\alpha)}}\sum_{r=1}^{\cal Z} \frac{(-1)^r}{r^\alpha}
    \label{Eq: hc2}
\end{equation}
which interestingly, depends on $\alpha$. Therefore, in sharp contrast to the short-range Ising model, few-neighbor Ising models are not symmetric against mirror inversion. In this paper, we mostly use $h_c^1 = 2$ as a point of comparison since it is the critical point for all the models including the SR Ising, the $ {\cal Z}$-neighbor extended Ising and the true LR models. Unlike critical points, the phase diagrams of the SR and LR models match as shown in Fig. \ref{fig:gap_closing}. The region between $h_c^1$ and $h_c^2$ belongs to the ordered phase, while the rest of the region is in the disordered phase. In the extreme limit of a high magnetic field, the system would be polarized in the $z$ direction, which clearly specifies the disordered (paramagnetic) phases. Note here that, unlike the transverse Ising model, the critical points as well as the phases are not symmetric across $h=0$.


Let us now move to the finite system, which also leads to the energy dispersion from Eq. (\ref{omegakLR}), for both the LR and the $\mathcal{Z}$-neighbor extended Ising models. 
As it can be readily seen from the plot of the spectrum in Fig. \ref{fig:dispersion},  near $k=0$, ~$\omega_k\sim k^{\alpha-1}$ when $1<\alpha<2$ for the true LR model. However, dispersion of the $ \cal Z$-neighbor extended Ising model is like the Ising one, $\omega_k\sim k$ when $k \approx 0$ for all $\mathcal{Z} \le 20$ which will be the main focus of this work. It implies that, in principle, only the true LR case supports the instantaneous information transfer since the fastest excited quasiparticle has infinite propagation velocity $v_g = d\omega_k/dk \sim k^{\alpha-2} \to \infty$, when $k \to 0$ for $1<\alpha<2$. Therefore, this result gives us the intuition that features of the ground state in the LR model can be distinct from those of the ground state of the few-neighbor Hamiltonian.  
This indicates that different patterns for bipartite entanglement between two sites of these two classes may emerge. However, we will show that it is still possible that the behavior of two-qubit entanglement of the true LR model will match that of the few-neighbor Hamiltonian. The appearance of such similar characteristics is possibly due to the fact that, in practice, the maximum velocity of the fastest quasiparticle is not infinity and it is bounded by the generalized Lieb-Robinson bound ~\cite{lieb_robinson_bound1972,PhysRevX.10.031009, PhysRevX.10.031010, PhysRevLett.123.250605, PhysRevA.104.062420}.

In the regime when $\alpha<1$, $h_c^1=2$ is still a critical point in the finite size system of the LR model and the dispersion looks like a $\delta$ function for the LR case (see Fig. \ref{fig:dispersion}). However, for the few-neighbor model, it is still an Ising-like dispersion but with different prefactors. 
This should also indicate that the true LR case is very different from the few-neighbor Hamiltonian.

\section{Outline for computing Correlations}
\label{sec:correlation}
From the diagonalization discussed in the preceding section, we are now ready to compute both classical and quantum correlations, in particular, two-qubit entanglement between two arbitrary lattice sites. Before studying the behavior of quantum correlation, let us describe the formalism used for computation.

\subsection{Classical correlation}
To compute bipartite reduced density matrices between any two sites, $i$ and $j$, we require an evaluation of magnetizations and long-range classical correlators. 
Suppose, the ground state of the system is $ |\wp_{0} \rangle$. The magnetizations at the site $i$ are defined as
\begin{equation}
\begin{aligned}
    m_{i}^{\alpha}=\left\langle\wp_{0}\left|\sigma_{i}^{\alpha}\right| \wp_{0}\right\rangle,
\end{aligned}
\end{equation}
with $\alpha = x, y, z$, while the correlation functions (correlators) between spins $i$ and $j$ can be represented as
\begin{align}
\mathcal{C}_{i j}^{\alpha \beta}&=\left\langle\wp_{0}\left|\sigma_{i}^{\alpha} \sigma_{j}^{\beta}\right| \wp_{0}\right\rangle,
\end{align}
where $\alpha, \beta = x,y,z$.
Since we are dealing with the ground state of a Hermitian Hamiltonian, the magnetization in the $y$ direction $m^y$ and correlators $\mathcal{C}^{x y}$, $\mathcal{C}^{y x}$, $\mathcal{C}^{y z}$, and $\mathcal{C}^{z y}$ vanish.  
To compute other correlators, let us define two operators,
\begin{equation}
\mathcal{A}_{i}=c_{i}^{\dagger}+c_{i}, \quad \mathcal{B}_{i}=c_{i}^{\dagger}-c_{i},
\end{equation}
and by using Jordan-Wigner transformations, magnetizations and classical correlators (CC) can be written in terms of $\mathcal{A}_{i}$ and $\mathcal{B}_{i}$ as
\begin{equation}
\begin{gathered}
m_{i}^{z}=-\left\langle\wp_{0}\left|\mathcal{A}_{i} \mathcal{B}_{i}\right| \wp_{0}\right\rangle, \\
\end{gathered}
\end{equation}
\begin{equation}
    \mathcal{C}^{xx}_{ij} = \langle \wp_{0} | \mathcal{B}_{i}\mathcal{A}_{i+1}\mathcal{B}_{i+1} \ldots \mathcal{B}_{j-1}\mathcal{A}_{j}| \wp_{0}\rangle,
\end{equation}
\begin{equation}
    \mathcal{C}^{yy}_{ij} = (-1)^{(j-i)}\langle \wp_{0} | \mathcal{A}_{i}\mathcal{B}_{i+1}\mathcal{A}_{i+1} \ldots \mathcal{A}_{j-1}\mathcal{B}_{j}| \wp_{0}\rangle,
\end{equation}
and \\ 
\begin{equation}
    \mathcal{C}^{zz}_{ij} = \langle \wp_{0} | \mathcal{A}_{i}\mathcal{B}_{i}\mathcal{A}_{j}\mathcal{B}_{j}| \wp_{0}\rangle.
\end{equation}
Here the magnetization $m^x$ and the correlation functions $\mathcal{C}^{xz} \text{ and } \mathcal{C}^{zx}$ vanish by means of Wick's theorem since these quantities involve an odd number of fermionic operators.  
To evaluate the rest of the correlators, we contract pairwise the product of operators again via Wick's theorem. Since the aforementioned operators, $\mathcal{A}_{i}$ and $\mathcal{B}_{i}$ obey anticommutation relations, only certain pairs give non-trivial values. Precisely, 
\begin{equation}
\begin{aligned}
\left\langle\mathcal{A}_{i} \mathcal{A}_{j}\right\rangle=\sum_{k} \phi_{k i} \phi_{k j} &=\delta_{i j},
\end{aligned}
\end{equation}
\begin{equation}
\begin{aligned}
\left\langle\mathcal{B}_{i} \mathcal{B}_{j}\right\rangle=-\sum_{k} \psi_{k i} \psi_{k j} &=-\delta_{i j},
\end{aligned}
\end{equation}
and
\begin{equation}
\begin{aligned}
\left\langle\mathcal{B}_{i} \mathcal{A}_{j}\right\rangle=-\left\langle\mathcal{A}_{j} \mathcal{B}_{i}\right\rangle=-\sum_{k} \psi_{k i} \phi_{k j} &=-\left(\boldsymbol{\psi}^{T} \boldsymbol{\phi}\right)_{i j}=\mathcal{G}_{i j}
\end{aligned}
\end{equation}
are the pairs that finally contribute to the expectation values. Here $\mathcal{G}$ is the correlation matrix which can be obtained from $\boldsymbol{\psi}$ and $\boldsymbol{\phi}$. In terms of $\mathcal{G}$, the non zero diagonal correlation functions read 

\begin{equation}
\mathcal{C}^{xx}_{ij} =
\begin{vmatrix}
\mathcal{G}_{i,i+1} & \mathcal{G}_{i,i+2} & \dots & \mathcal{G}_{i,j} \\ 
\vdots \\
\mathcal{G}_{j-1,i+1} & \dots & \dots & \mathcal{G}_{j-1,j}
\end{vmatrix},
\end{equation}

\begin{equation}
\mathcal{C}^{yy}_{ij} =
\begin{vmatrix}
\mathcal{G}_{i+1,i} & \mathcal{G}_{i+1,i+1} & \dots & \mathcal{G}_{i+1,j-1} \\ 
\vdots \\
\mathcal{G}_{j,i} & \dots & \dots & \mathcal{G}_{j,j-1}
\end{vmatrix},
\end{equation}
and
\begin{equation}
\mathcal{C}^{zz}_{ij} =  (\mathcal{G}_{ii}\mathcal{G}_{jj} - \mathcal{G}_{ji}\mathcal{G}_{ij}).
\end{equation}
By solving Eqs. (\ref{A_B}) and (\ref{bogo}), we can compute magnetization and all the CCs.

\begin{figure}[h]
    \centering
   \includegraphics[width=9cm,height=11.5cm]{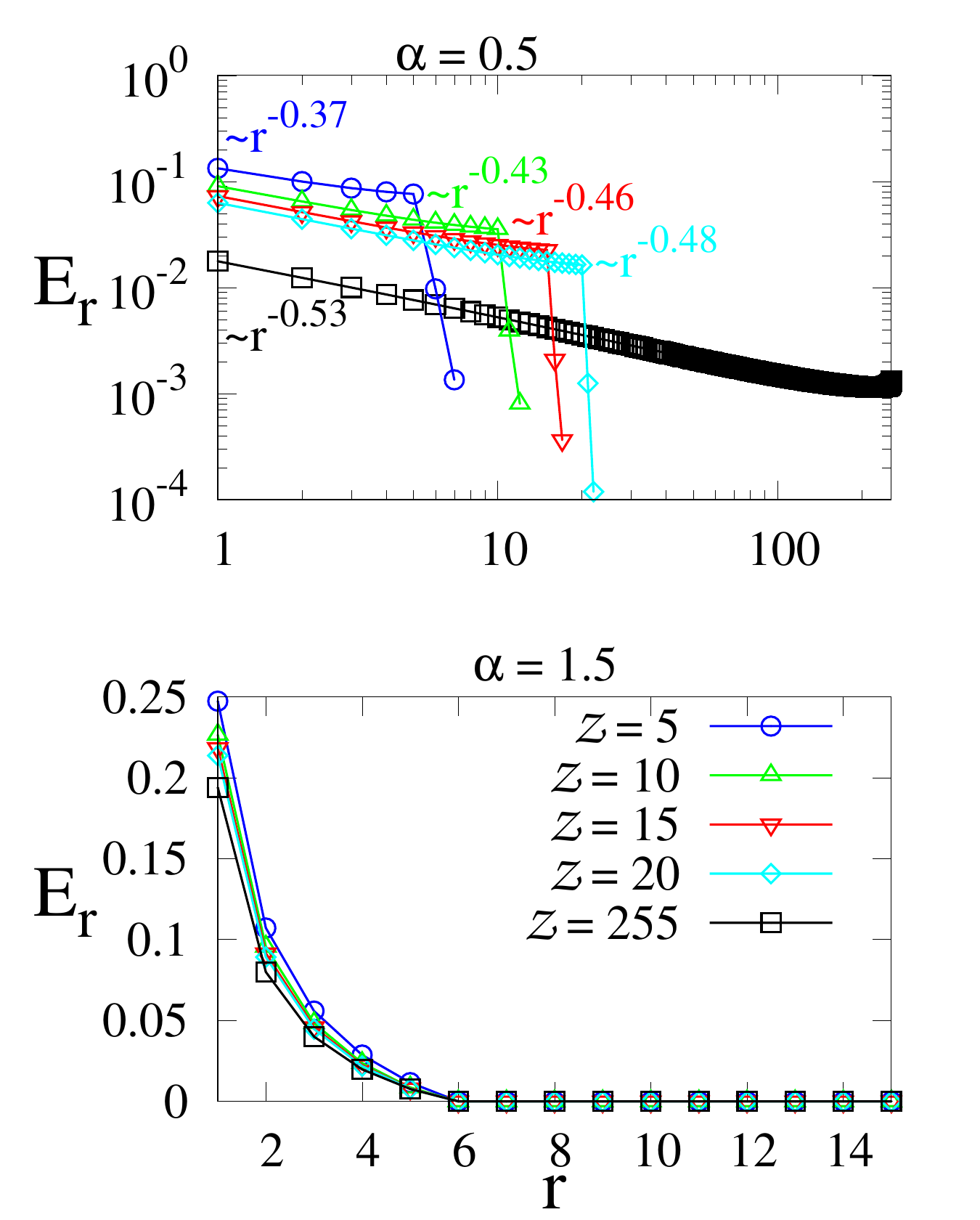}
    \caption{(Color Online.) \textbf{Entanglement pattern in quasi-local and non-local regimes:} Variation of $E_{r}$ (ordinate) as a function of $r$ (abscissa). Different symbols represent different coordination number, $\mathcal{Z}$. Note that $\mathcal{Z} = 255$ indicates the fully-connected model. Here $h = 2.5$. In the quasi-local regime (lower panel), the variation of $E_r$ with $r$ follows the law governed by $r^{\beta}$, where $\beta = \{-0.37,-0.43,-0.46,-0.48,-0.53\}$, such that $ \beta \approx \alpha$ for the corresponding $\mathcal{Z} = \{5,10,15,20,255\}$. Both the axes are dimensionless.}
    \label{fig:lr_Er_r}
    
    \end{figure}
    
\begin{figure}[h]
    \centering
   \includegraphics[width=9cm, height= 6.5cm]{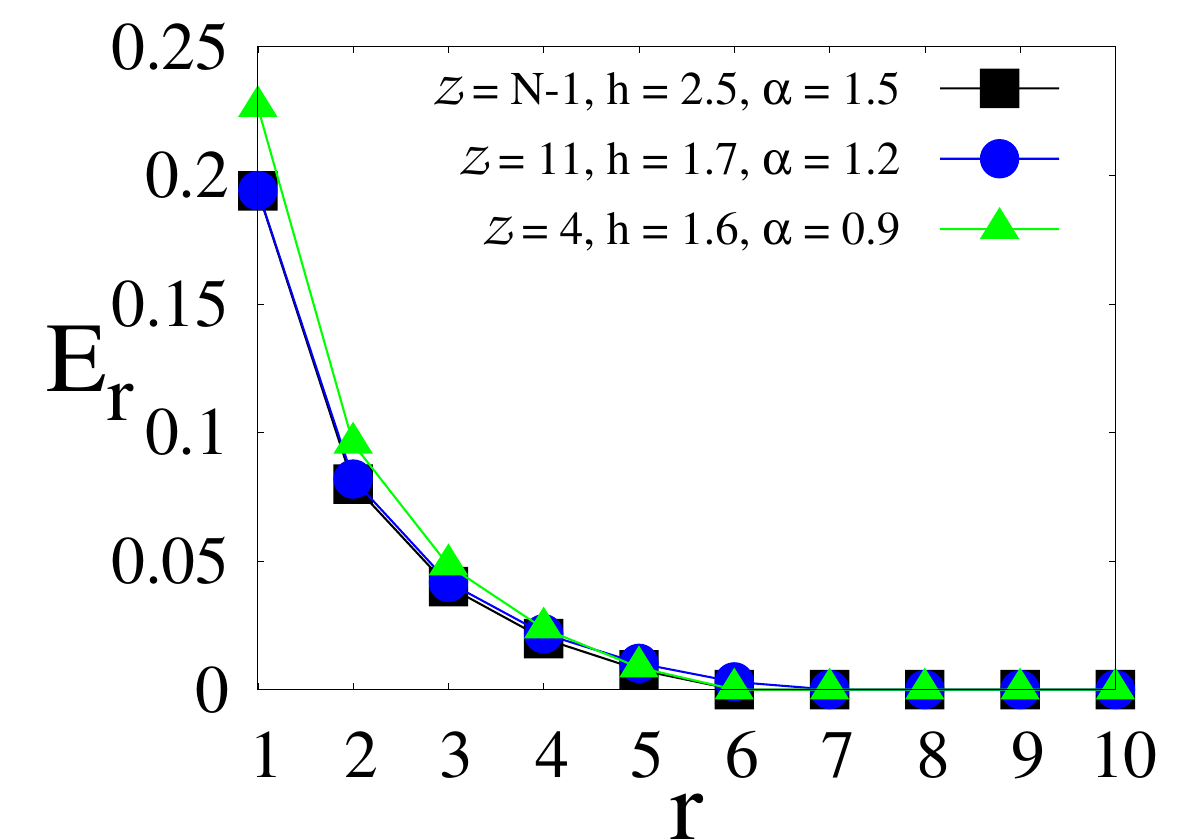}
    \caption{(Color Online.) \textbf{Entanglement trends of a fully connected model, mimicking a few-neighbor model.} Two-qubit entanglement $E_{r}$ ($y$-axis) is plotted vs the variation of $r$  ($x$-axis) for different values of $(\mathcal{Z}, h, \alpha)$. For a given \(r\) value, equal values of entanglement for different sets of values are observed which demonstrates that a few-neighbor model can mimic the entanglement of the ground state in the LR model efficiently by adjusting the system parameters accordingly. Both axes are dimensionless.}
    \label{fig:Er_r_mimick}
\end{figure}



 \subsection{Quantum correlation}

We are interested in investigating the trends of pairwise entanglement between two lattice sites $i$ and $j$ in the ground state of the Hamiltonian. From the nonvanishing transverse magnetization and classical correlators, the two-party reduced density matrix obtained from the ground state becomes 
\begin{eqnarray}
 {\rho}_{ij}&=&\frac{1}{4}\Big[\mathbb{I}_i\otimes\mathbb{I}_j
 +m^z_i\sigma^z_i\otimes\mathbb{I}_j+\mathbb{I}_i\otimes m^z_j\sigma^z_j
 \nonumber\\
 &&
 +\sum_{\alpha=x,y,z}\mathcal{C}^{\alpha\alpha}_{ij}\sigma^\alpha_i\otimes\sigma^\alpha_j
 \Big].
\label{rho_fermi_equiv} 
\end{eqnarray}
We can immediately determine any quantum correlation measure, especially an entanglement measure which is a nonlinear function of $m_{z}$ and $\mathcal{C}^{\alpha \alpha}$. In this work, we compute logarithmic negativity \cite{neg4,neg5} for investigation. Apart from bipartite entanglement, we are also interested in examining the distribution of entanglement of the ground state among different sites, quantified via monogamy of entanglement \cite{coffman2000,monorev,ckw2}.


\section{Ground state entanglement in LR and few-neighbor models }
\label{sec:gs_ent}

  Long-range models are known to have rich characteristics which are typically not present in the nearest-neighbor model. Hence the LR or a few-neighbor model requires a careful analysis from the scratch. For example, even without the absence of scale invariance at a quantum critical point, the classical correlations of an LR model are allowed to have an infinite correlation length, thereby spreading over the entire system~\cite{Vodola2,Cevolani2016,LengthInLR}. 
  
  On the other hand, quantum correlation, especially entanglement is known to be fragile compared to classical correlations and cannot be shared arbitrarily between different parts of the systems due to the monogamy property~\cite{coffman2000}. This in turn should restrict entanglement from having an algebraic scaling or an infinite entanglement length. 
  
  Until recently, most of the studies on entanglement have been restricted to one-dimensional nearest-neighbor quantum spin models. The primary reason behind such investigation is the existence of a method by which one can compute several features analytically both for finite system size and in the thermodynamic limit. Moreover, with the advent of tensor networks in the past decade, a variety of numerical techniques have been developed which make LR models tractable with good enough accuracy~\cite{PhysRevLett.109.267203,VUMPS,1801.00769,Laurent2018, Laurent2021, Laurent2021, Hauke2021,prx1, prb1, prb2, pra1}. For example, the entanglement area law typically holds for SR systems in one dimension~\cite{0705.2024,hep-th/0405152,quant-ph/0503219} which is not guaranteed in LR systems, although the success of tensor network-based numerical techniques in quasi-local regimes of the LR systems suggests that at least in those regimes the area law is not strongly violated.

  The twin restrictions of entanglement area law and monogamy of entanglement hinder the spreading of entanglement in true LR systems. In a quantum network, LR systems are typically used as the underlying architecture, although preparing a true LR model can be immensely costly as well as difficult in some physically realizable systems due to the increment of noise in the system. Hence the question arises whether it is worth developing such an LR system which leads to a reasonable spread (distribution) of entanglement. In general, in a digital quantum computer e.g.,  quantum approximate optimization algorithm, simulation on superconducting circuits, as mentioned before, we require ${O}(N!)$ two-qubit gates to implement a true LR system of system size $N$. The question that we address here is the following: Can one achieve the same distribution of entanglement using only a few two-qubit gates? Therefore, one could significantly reduce the noise if an exponential-to-algebraic reduction of two-qubit gates can be achieved. Such intuition has already been implemented in the chimera setup in the D-wave quantum annealing computers~\cite{Boothby2015} to mimic the LMG model ($\alpha=0$ case)~\cite{LMG} with a limited number of interacting qubits.  
 
In the Hamiltonian considered here, we have two tuning parameters that can control the long-range interactions of the model. The first one is the exponent $\alpha$; if we move from $\alpha = \infty$ to $\alpha=0$, we continuously go from the nearest-neighbor Ising to the end-to-end connected LMG model where all pairs of interaction have the same strength, independent of the distance between the pair of spins. Note that, except when $\alpha = \infty$, the number of two-qubit gates required in all other cases is the same (exponential with $N$). The other parameter that we can regulate to achieve the same control is to manually increase the number of pairing interactions from $2$ to $\infty$ in the thermodynamic limit. For a finite system of size $N$, when $\mathcal{Z} = 1$, we get the NN Ising model, while when ${\cal Z}=N-1$, we have the end-to-end extended Ising model with open boundary conditions. The first question that we answer here is whether for the same algebraically decaying interaction ($\alpha$ is the same in both models) it is possible to mimic the behavior of two-qubit entanglement of the fully connected model with a few interactions.

\emph{Mimicking the true LR model with $\alpha > 1$. } To answer the above question, we first study the pattern of two-qubit entanglement as a function of the distance for different values of ${\cal Z}$.
When $\alpha > 1$, we find the answer to be affirmative. In other words, in classes of quasi-local models, the two-qubit entanglement of a finite-range system indeed behaves like the true LR model. Specifically, entanglement between two arbitrary sites, $i$ and $j$, denoted by $E_{r}$, with $r = |i - j|$ the distance between sites $i$ and $j$ (as depicted in Fig. \ref{fig:lr_Er_r}), decreases with the increase of $r$ and finally vanishes at $r=r_c$, above which $E_r$ is zero even for the true LR model. 
If we now compare $E_{r}$ obtained from the model with $\mathcal{Z} \ll N$, we find that indeed $E_{r} \to 0$ as $r \to r_{c}$ when $\mathcal{Z} \backsim \mathcal{Z}_{c}$, thereby simulating the equivalent feature of the LR model by a $\mathcal{Z}_{c}$-neighbor extended Ising model. With a decreasing value of $\alpha$, $r_c$ increases for the LR model. We observe that the number of required pairing interactions ($\mathcal{Z}_{c}$) also increases. After a careful numerical search, we conclude that $\mathcal {Z}<O(10^1)$ when $1<\alpha<2$, i.e., the quasi-local regime.

\emph{No resemblance of the $\mathcal{Z}$-neighbor model to the LR model having $\alpha < 1$. } If we move towards the deep non-local regime ($\alpha \leq 0.6$ for $N = 256$), entanglement becomes long-range with an algebraic scaling roughly as $ E_r \approx r^{-\beta}$, where $\beta \simeq \alpha $ and therefore, $r_{c} \sim O(N)$. To mimic the long-range entanglement, the minimum number of finite-range interactions (${\cal Z}_c$) approaches $O(N)$. This implies that the entanglement $E_{r}$ of the LR model can no longer be simulated by a finite number of interactions.  We show in Fig. \ref{fig:lr_Er_r} (upper panel), that no ${\cal Z} \neq N-1$ is sufficient to mimic the entanglement in the LR model.  On the other hand, close to the transition between quasi- and non-local regimes, $E_r$ vanishes at some finite $r_c \ne N-1$ which implies that the model can be mimicked by finite $Z_c \approx O(r_c)$, even when $\alpha < 1$. However, $Z_c$ is higher in this situation compared to the $\mathcal{Z}_c$ observed in the quasi-local regime.

Up to now, when comparing the different classes of models, we always chose the same value of $\alpha$ for all the considered ${{\cal Z}}$ values. At this point, let us ask another reasonable question: Can we further reduce ${\cal Z}_c$ if we can also tune the value of $\alpha$ for the finite-neighbor Hamiltonian $?$ We observe that the value of ${{\cal Z}}$ has a monotonic relationship with $\alpha$, i.e., the trends of entanglement of a true LR model with exponent $\alpha$ can have remarkable resemblance to a fewer pairing interaction (i.e., $\mathcal{Z}<{\cal Z}_c$) having exponent $\alpha' < \alpha$ (see Fig. \ref{fig:Er_r_mimick}).

In Fig. \ref{fig:lr_Er_r} we choose the magnetic field as $h=2.5$, which belongs to the disordered phase. In the following section, we report the behavior of entanglement in the different phases of both the LR and the few-neighbor models. 

  \begin{figure}[t!]
    \centering
    \includegraphics[width=8.5cm,height=11.5cm]{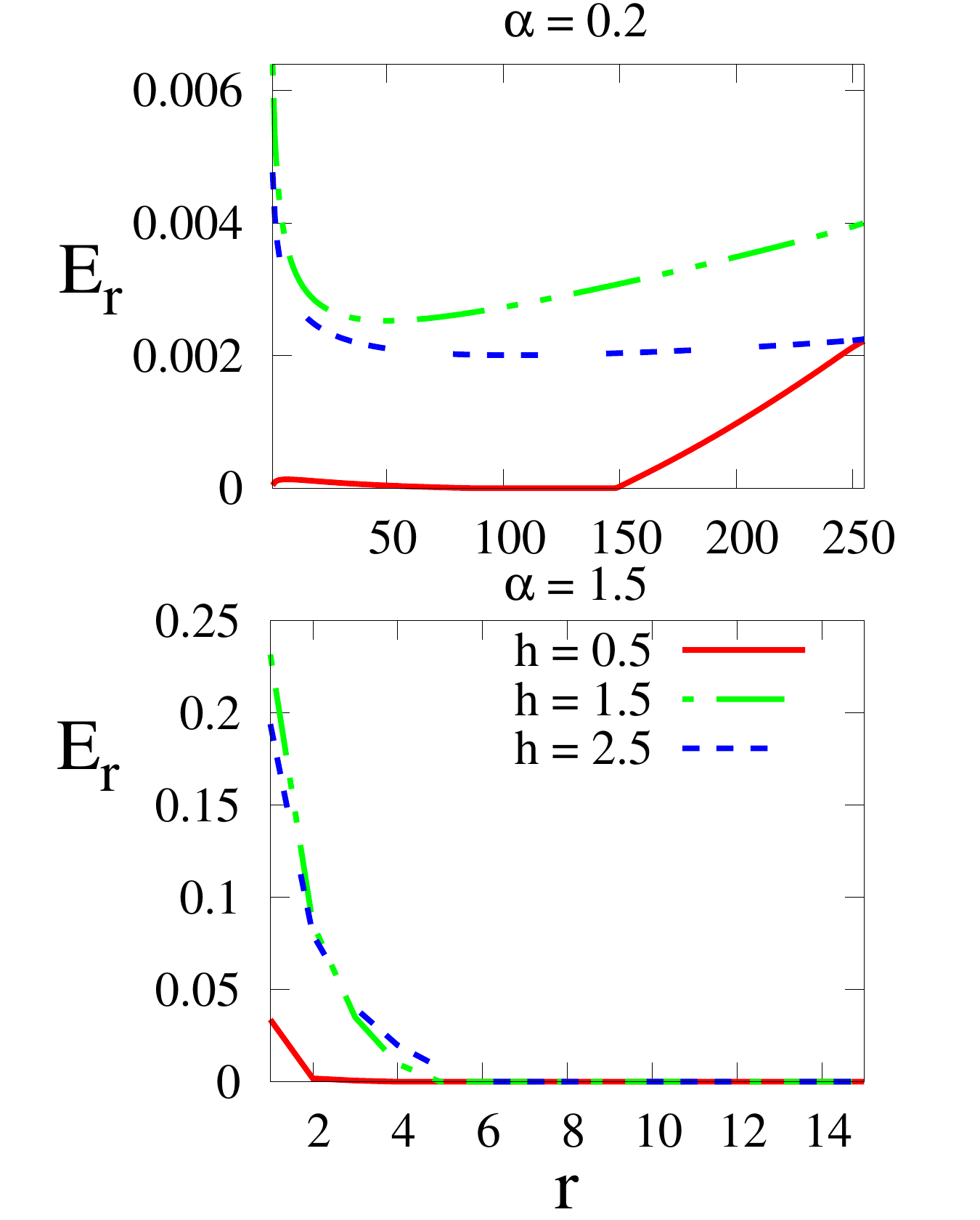}
    \caption{(Color online) \textbf{Effects of phases on entanglement}.  $E_r$ (ordinate) against $r$ (abscissa). Solid, dashed and dotted lines correspond to different values of $h$ while upper and lower panels are for non-local and quasi-local regimes respectively. The model is fully connected, i.e., $\mathcal{Z} = N-1$. Counter-intuitively,  for $\alpha<1$ (upper panel), pairwise entanglement is higher for spins separated by longer distance $r$ compared to that between nearby sites. Both axes are dimensionless.}
    \label{fig:Er_r_alpha}
\end{figure}

\section{Entanglement in different phases of the models}
\label{sec:ent_phase}
We will now investigate the effects of the magnetic field on the pairwise entanglement, thereby changing the phases of the system along with the change of interaction strength and number of interacting pairs. In general, in the disordered phase, the value of entanglement decreases with the increase of $h$, i.e., when we move deep into the disordered phase. However, due to the monogamy property of entanglement, such a decreasing pairwise entanglement also has some beneficial role in the system. Specifically, the spread of entanglement, i.e., the number of non-vanishing pairwise entanglement between two sites $i$ and $j$, increases with the increase of $h$. However, at $h\to \infty$, the ground state is fully separable, and therefore both bipartite and multipartite entanglement vanish, which also implies that individual terms in the monogamy score also goes to zero, thereby leading to a vanishing monogamy score. This suggests that there is a trade-off between the pairwise entanglement and the distribution of entanglement in the system. 

The contrasting entanglement spread over pairs of distant neighbors that we will report now in different phases of the system can only be seen in the true LR model, not in the $\mathcal{Z}$- neighbor extended model. Hence we concentrate on the patterns of entanglement in the disordered and the ordered phases of the true LR system i.e., the model having $\mathcal{Z} = N-1$. 
  \begin{figure}[h]
    \centering
   \includegraphics[width=8cm, height= 6.2cm]{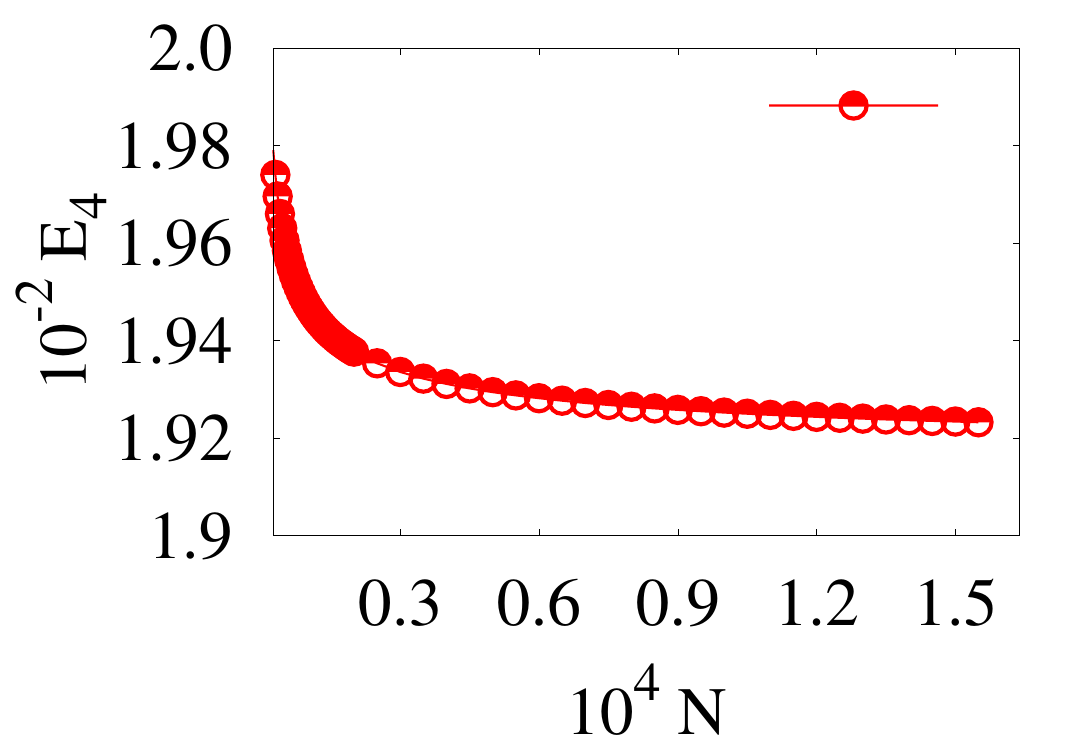}
    \caption{\textbf{ Entanglement in the thermodynamic limit:}. Entanglement between first and fifth spins, $E_4$ (ordinate) with respect to the system size $N$ (abscissa). Here $\alpha = 1.5$ is chosen in the quasi-local regime and $h = 2.5$.  Both axes are dimensionless.}
    \label{fig:Er_r_N}
\end{figure}
As discussed in Sec.~\ref{sec:critical}, the quantum phase transition point,  which is common to both the LR and the ${\cal Z}$-neighbor systems, is at $h_{c}^{1} = 2$. Using the same analogy known for the SR Ising model \cite{qptbook2}, at the two extremum points at $h=0$ and $h=\infty$, the ground states are product, and therefore entanglement vanishes at both points.   
Usually, as we move from the deep disordered phase at $h=\infty$ towards the critical point at $h_{c}^{1}=2$, we expect that the pairwise entanglement increases as shown in Fig.\ \ref{fig:Er_r_alpha}. 

Let us first concentrate on the non-local regime, i.e., $\alpha < 1$. In the disordered phase i.e., when $h$ is high enough, bipartite entanglement between different neighbors, $E_r$ first decreases and then saturates with the variation of $r$ (see Fig.\ \ref{fig:Er_r_alpha}). If we move towards the critical point, $h_c^1 = 2$, the pairwise entanglement content for a given $r$ increases. Surprisingly, entanglement between distant pairs also increases with the increase of $r$, resulting in a U-shaped entanglement pattern as a function of $r$. In general, it is expected that the bipartite entanglement between spins decreases when the distance between spins $r$ increases. However, such an intuition does not hold for $\alpha < 1$, e.g., we observe that, $E_r$ increases with $r$ after a certain $r$ value in both ordered and disordered phases.

The U-shaped behavior of entanglement can be explained in terms of the boundary effect of the spin chain. We characterize the effect of the boundary by considering the left-most spin as the nodal spin and calculate the entanglement between the nodal spin and the spin at a distance $r$ from the nodal spin.
Now, as we increase $r$, the number of neighbors with comparatively strong interaction strength (when $\alpha>0$) increases until the middle of the spin and thereafter decreases as the non-nodal spin approaches the right-most spin. Therefore, the spin pair $(1, \frac {N}{2})$ has more strongly interacting neighbors than the spin-pair $(i, N)$ at the boundary. By virtue of the monogamy constraint, it is, therefore, expected that the $(1, \frac {N}{2})$ pair would have less entanglement as they have more strongly interacting neighbors as compared to the $(i,N)$ pair.

Here, the boundary effect plays a positive role on both sides, as opposed to spins situated in the middle. Unlike the boundary spins, for the spins that are situated in the middle of the chain, the entanglement is restricted by the monogamy constraint in such a way that the amount of entanglement shared on either side have to be equal, i.e., $ E_{\frac{N}{2}:\frac{N}{2}+i}  = E_{\frac{N}{2}:\frac{N}{2}-i}, \forall \text{  }i \in \{1,\frac{N}{2}-1\}$. This effect can be attributed to the higher entanglement behavior at the right boundary. Another positive effect contributing to the generation of more entanglement is the large number of $\sigma_z$ operators present in between the nodal spin and the spins that are situated at the extreme right. Such an effect is absent in the left boundary; thus the amount of entanglement is less than at the right boundary, although higher compared to the middle spins. Also, in the presence of a smaller magnetic field strength, $h$, the effect of the interaction is more prominent.

Notice, however, that such a behavior is not universal for $\alpha < 1$ and it depends on $N$. In particular, with the increase of $N$, the value of $\alpha$ for which such a distant neighbor entanglement is created also changes.  
As it has already been argued in the previous section, the model cannot be simulated with a finite range of the interacting model. This is in sharp contrast with the previous results~\cite{VodolaThesis, Maity_2019, PhysRevA.97.062301,canovi2014} and cannot simply be explained by the violation of the entanglement area law. However, crossing the critical point $h_{c}^{1} = 2$, if one moves towards the product state at $h=0$, the value of entanglement reduces further which is illustrated by the red solid line ($h=0.5$) in Fig. \ref{fig:Er_r_alpha} (upper panel). The $U$-shaped pattern, however, persists in the ordered phase. 

Let us now deal with the quasi-local regime ($1<\alpha<2$). Entanglement is always short-ranged here and vanishes after a certain $r$. Therefore, entanglement in this regime has the typical expected behavior, i.e., entanglement decays with $r$ when $\alpha > 1$. 
In most cases, for any $r>{O}(10^1)$, $E_r$ becomes zero and, therefore, no $U$-shaped pattern is observed in the quasi-local regime. The other features across different phases of the LR model remain the same. Specifically, deep in the ordered phase, i.e., in the neighborhood of $h = 0$, entanglement decreases and at the same time, becomes short-ranged. For example, for $h=0.5$ we observe that the entanglement survives only up to the next-nearest-neighbor even though we are dealing with a fully connected LR model (as depicted in Fig. \ref{fig:Er_r_alpha} (lower panel)). On the other hand, with the increase of $h$, especially, in the vicinity of $h_{c}^{1} = 2$, the value of pairwise entanglement is substantial and it survives for certain but low values of $r$.


%

  

\emph{Scaling. } It is interesting to determine whether the results described also holds with the increase of system size. We find that the results remain valid with the increase of $N$. Although the entanglement value decreases with $N$, the decrement slows down as $N$ increases and $E_r$ saturates when $N > 10^3$ (see Fig. \ref{fig:Er_r_N}). It manifests that the trend of entanglement length for moderate values of $N$ mimics the entanglement behavior that can be expected in the thermodynamic limit.

\begin{figure}[b]
    \centering
   \includegraphics[width=9cm,height=11.5cm]{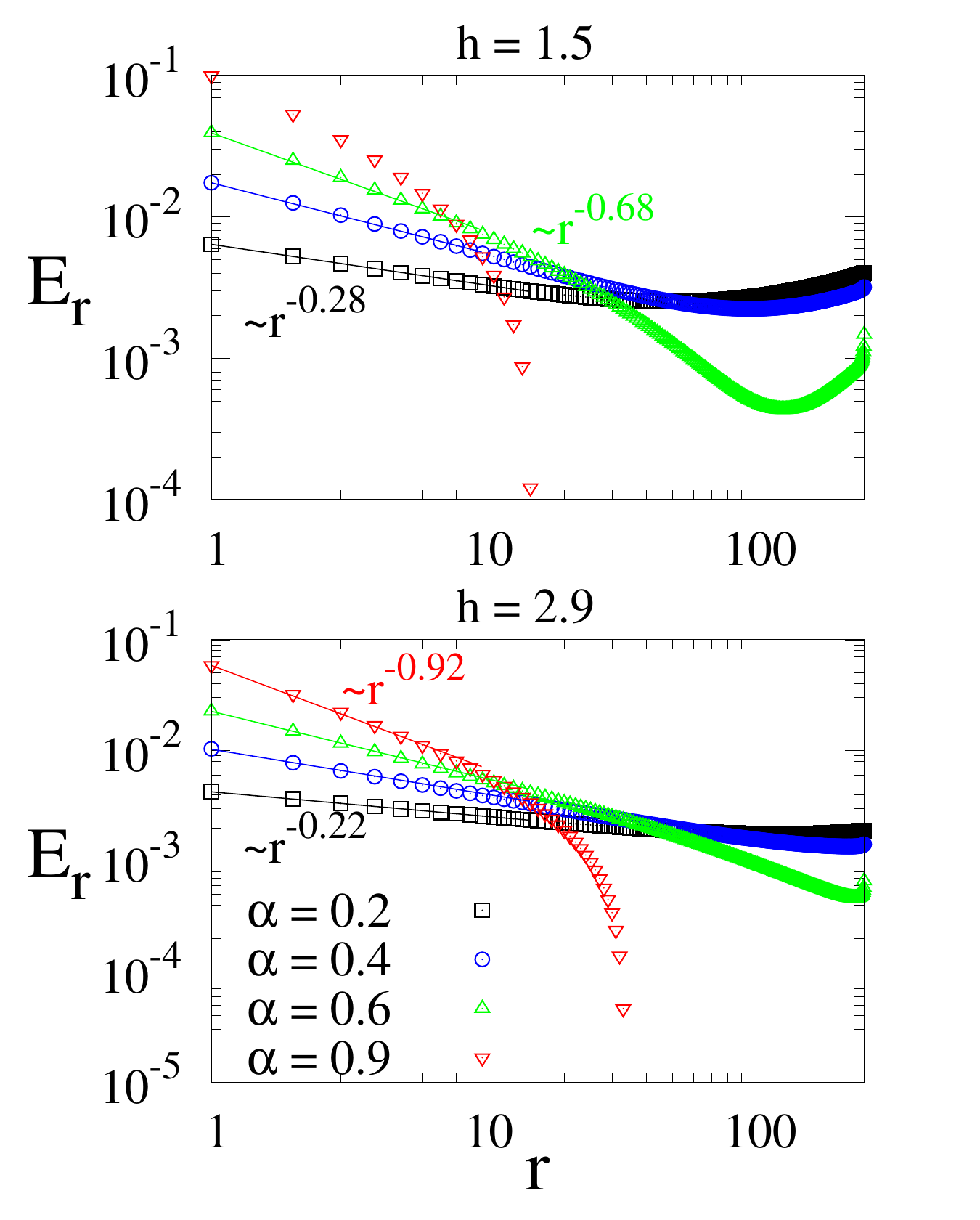}
    \caption{(Color Online.) \textbf{Functional dependence of $E_r$ on $r$ in the non-local regime and disordered phase.} Variation of $E_{r}$ (vertical axis) is plotted vs $r$ (horizontal axis) for different values of $\alpha$ in a fully connected model i.e., $\mathcal{Z} = N-1$. The upper and lower panels demonstrate two different phases, ordered and disordered phases respectively of the system, both depicting the same features.  The solid lines are the numerical fits for the corresponding $\alpha$ values. For a small range of $r$, entanglement scales as  $r^{-\beta} $, where $\beta \approx \alpha$ in both the phases except near the transition regime $\alpha\sim 1$, where entanglement is short-ranged.  In particular, the numerical fits are $\{r^{-0.28},r^{-0.49},r^{-0.68}\}$ (upper panel) and $\{r^{-0.22},r^{-0.42}, r^{-0.62},r^{-0.92}\}$ (lower panel).  Both the axes are dimensionless.}
    \label{fig:Er_r_vary_alpha}
\end{figure}
\subsection{Quasi-local vs non-local regime: Entanglement behavior}

To make the comparison between entanglements in quasi-local and non-local regimes, we consider different $E_r$ for both $\alpha > 1$ and $\alpha < 1$. Depending on the tuning parameter $\alpha$ which dictates the strength of interactions between neighbors, we determine contrasting behavior in entanglement. In particular, in the local regime, unlike classical correlations, entanglement is always short-ranged in both phases. Therefore entanglement in the LR model can always be mimicked with only interactions between a few-neighbors. On the other hand, different behavior emerges in the non-local regime, i.e., when $\alpha < 1$.


When $\alpha\approx 1$, i.e., when the system is at the cross-over between the quasi-local and non-local regimes, entanglement remains short-ranged, i.e., only a few $E_r$ remain nonvanishing. As we move towards the deep non-local regime, entanglement becomes fully connected, and U-shaped, i.e., non-monotonic with $r$ in both the disordered and ordered phases. Although with the further reduction of the $\alpha$ value towards the LMG model with uniform interaction strength~\cite{Morigi_2018}, the U-shaped pattern of $E_r$ with $r$ gets flattened, the counter-intuitive nonmonotonic behavior of entanglement with $r$ is more prominent in the ordered phase compared to the disordered phase. More importantly, the entanglement in the disordered phase has an algebraic tail with increasing $r$ if we neglect the few farthest spins (the part after it becomes minimum). To be precise, in the disordered phase, i.e.,  when $\alpha<1$, $E_r$ scales as $\sim 1/r^{\alpha}$ $\forall ~ \alpha$ where $E_r$ is decreasing with increasing $r$ (see Fig. \ref{fig:Er_r_vary_alpha}).

In the non-local regime of the LR model, it is possible to have end-to-end connected entanglement and hence a natural question to address is to determine the distribution of entanglement between different pairs. For example, if the ground state possesses an end-to-end entanglement but with a vanishing value, the entanglement may not be so useful for a multiparty quantum information processing task. Therefore, we now look for scenarios where the monogamy bound is saturated which can be though of an optimal spread of entanglement for a given quantum information protocols.  

\begin{figure}[t]
    \centering
    \includegraphics[width=9cm,height=12.5cm]{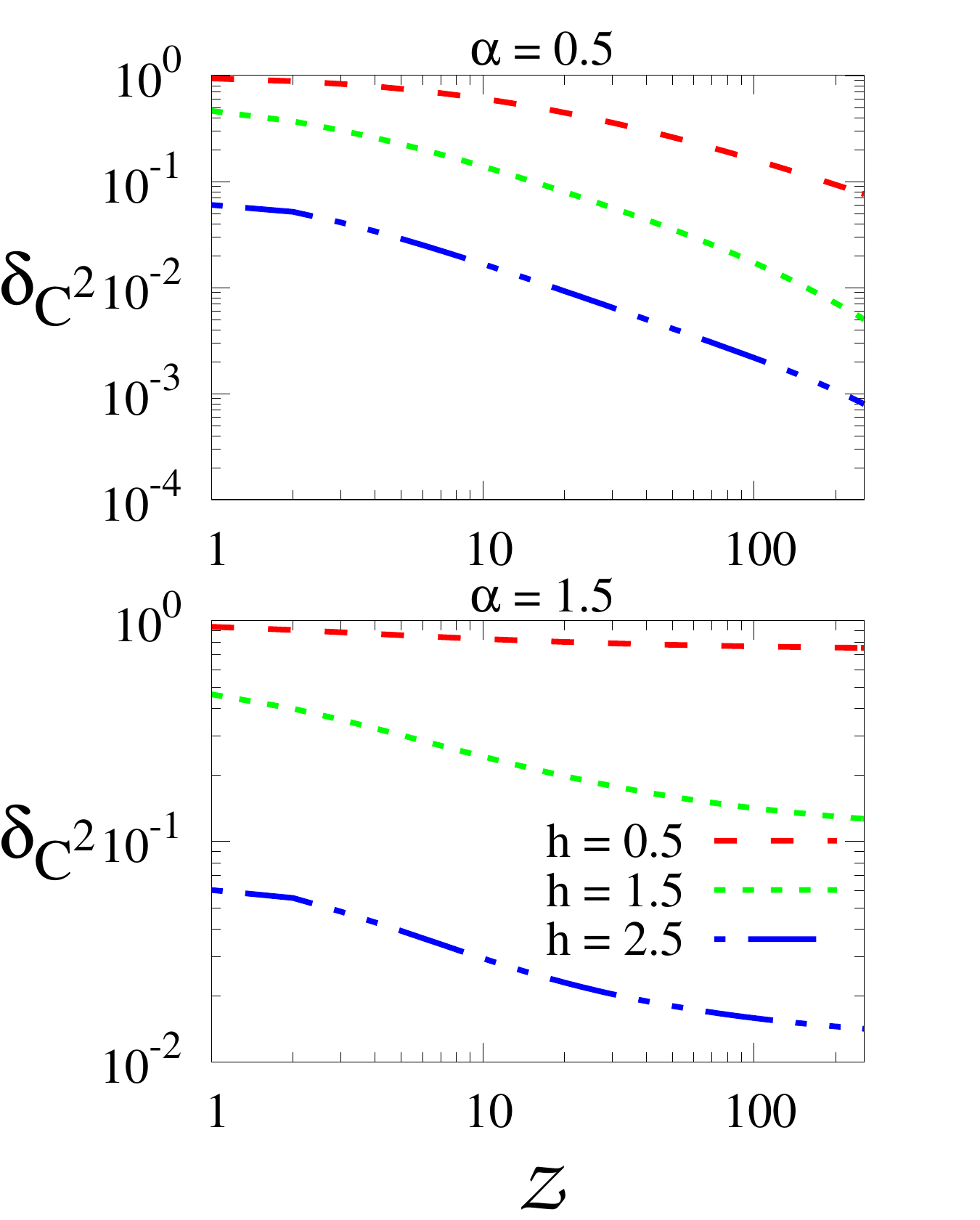}
    \caption{(Color Online.) \textbf{Monogamy score as an indicator for distribution of entanglement.}  Monogamy score ($\delta_{C^{2}}$) (ordinate) with $\mathcal{Z}$ (abscissa). Different lines indicate different values of $h$ in the non-local (upper panel) and the quasi-local (lower panel) regimes. Both the axes are dimensionless.}
    \label{fig:M_a_05_15}
\end{figure}

\subsection{Contrasting characteristics of monogamy scores in different phases of the model}
\label{sec:monogamy}

 To capture the spread of entanglement among the pairs, we examine the entanglement monogamy score. We are interested in the scenario where, $\delta_{C^{2}} = 0$, i.e., $C^2_{1, \text{rest}} = \sum_{i=2}^N C^2_{1i}$.

 As we have seen before, entanglement is short-ranged in the quasi-local regime and, therefore, we can expect that the monogamy score will be far from vanishing. However, in the non-local regime, entanglement is long-ranged and can provide a bound on the distribution of the entanglement at the thermodynamic limit. For example, near $\alpha \approx 0$, i.e., for the LMG model, pairwise entanglement is mostly flat with $r$ in the disordered phase. 

\begin{figure}[t]
    \centering
    \includegraphics[width=8.5cm,height=12.5cm]{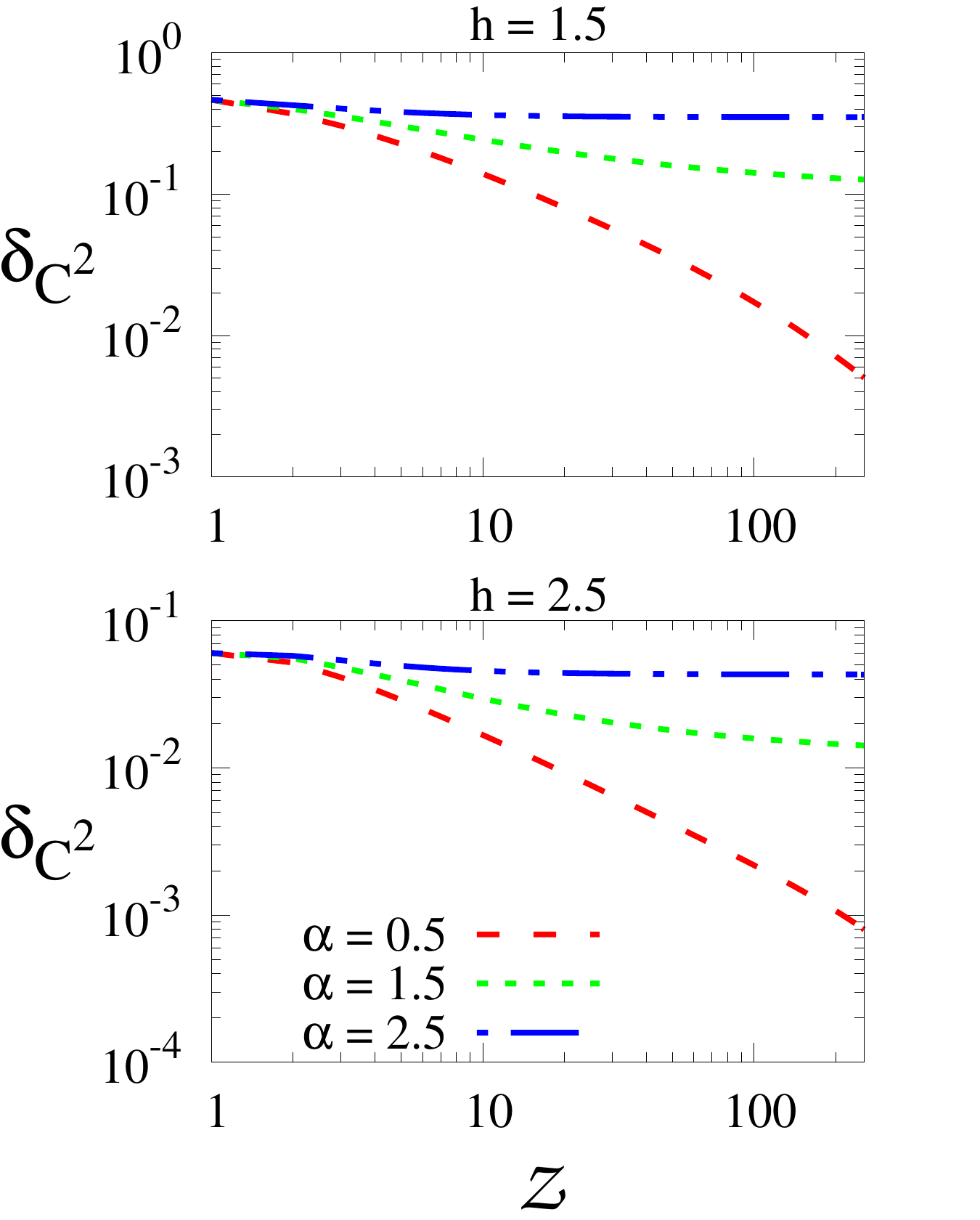}
    \caption{(Color Online.) \textbf{ Monogamy score with coordination number.} $\delta_{C^{2}}$ (vertical axis) vs $\mathcal{Z}$ (horizontal axis). Different kinds of lines represent different values of $\alpha$. Lower panel is for $h = 2.5$ while upper one indicates $h = 1.5$, i.e., two different phases of the system. Both axes are dimensionless.} 
    \label{fig:mono_Z_alpha}
\end{figure}




In general, we expect that the value of entanglement should decrease as we go deep into the discorded phase. However, we also expect that entanglement gets well distributed throughout the spin chain in this phase, as depicted in Fig. \ref{fig:M_a_05_15}. In particular, the monogamy score decreases as one increases $h$, thereby moving toward the disordered phase. If we compare the same with Fig. \ref{fig:Er_r_alpha}, we notice that the value of entanglement also decreases in the disordered phase. 

On the other hand, in the quasi-local regime (Fig. \ref{fig:M_a_05_15}  (lower panel)) i.e., when $1<\alpha<2$, the monogamy score does not have an algebraic tail but instead saturates to a constant value with the increase of ${\cal Z}$, although overall pattern of $\delta_{C^{2}}$ across different phases remains the same. The saturation of the monogamy score indicates that such LR models can be simulated with a model having finite-neighbor interactions. However,   it is difficult to numerically evaluate the optimal ${\cal Z}_c$ from the monogamy score which can mimic the true LR model, since there is no sharp change in the pattern of the monogamy score.





To monitor the dependence of the monogamy score on the phases along with the $\alpha$, we consider three different regimes of $\alpha$ and two different values of $h$, belonging to ordered and disordered phases. When $\alpha>2$, i.e., in the Ising universality class, the monogamy score is flat (blue dot-dashed line in Fig. \ref{fig:mono_Z_alpha}) with increasing ${\cal Z}$ which implies that the overall behavior of entanglement is similar to the SR model and, therefore, ${\cal Z}_c \sim {O}(1)$ should be enough to mimic the true LR model. In the intermediate quasi-local regime, i.e., $1<\alpha<2$, we find that the monogamy score decays with the increase of ${\cal Z}$, although its inclination changes to a shallow decay and ultimately becomes flat with $\mathcal{Z}$. It is in good agreement with the previous finding that $\mathcal{Z}_c \sim O(10^1)$ is enough to mimic the fully connected LR model. However, as $\alpha < 1$ (except for the transition from quasi-local to non-local regimes), $\delta_{C^{2}}$ has an algebraic tail, which illustrates that there is no ${\cal Z}_c \neq N-1$ which can behave similarly to the true LR system.


In summary, we point out that the monogamy score is a good indicator of entanglement distribution in the system. The saturation of the monogamy score for a finite $\mathcal{Z}$, which happens only when $\alpha > 1$, indicates the similar result that the fully connected model can be simulated only with a few pairing interactions.


\section{Conclusion}
\label{sec:conclu}

Among available physically realizable systems, long-range interactions arise naturally in some systems such as trapped ions, while there exist systems in which realizing LR models is costly. From both theoretical and experimental points of view, simulating the LR model is an important task to understand many exotic properties responsible for counter-intuitive phenomena that are typically absent in the corresponding short-range models. 

In this work, we showed that to examine the behavior of entanglement between any pairs in the ground state, it is not necessary to consider a fully connected model. In particular, we demonstrated that a model having few neighbor connections is sufficient to faithfully mimic the behavior of entanglement in the ground state of a fully connected model. However, we showed that such resemblance is not ubiquitous; it depends on the falloff rates of interactions, denoted by $\alpha$. Specifically, patterns of two-party entanglement of the LR model match with the model of few-range interactions only when $1< \alpha < 2$, which we call the quasi-local regime. Counterintuitively, when $\alpha < 1$, we observed that entanglement between the spins that are separated by a longer distance is higher than those pairs that are spatially closer to each other. Moreover, in this model, we reported that in the quasi-local regime, as the amount of external magnetic field increases, the amount of entanglement between spins decreases although the range of entanglement is strikingly increasing. Considering monogamy of entanglement, we illustrated that in the quasi-local regime, the monogamy score for entanglement saturates with the range of interactions, thereby demonstrating that a few range of interactions is enough to mimic entanglement in the LR system. On the other hand, the monogamy score of the LR system whose entanglement can not be reproduced by a few range of interactions, vanishes with the range of interactions, which is in good agreement with the results obtained for pairwise entanglement.

\acknowledgements

LGCL, SG, and ASD acknowledge the support from the Interdisciplinary Cyber Physical Systems (ICPS) program of the Department of Science and Technology (DST), India, Grant No.: DST/ICPS/QuST/Theme- 1/2019/23. 
DS is supported
by the grant 'Innovate UK Commercialising Quantum
Technologies' (application number: 44167). 
We  acknowledge the use of \href{https://github.com/titaschanda/QIClib}{QIClib} -- a modern C++ library for general purpose quantum information processing and quantum computing (\url{https://titaschanda.github.io/QIClib}), and the cluster computing facility at the Harish-Chandra Research Institute.

\bibliography{bib}
\appendix
 

\section{Logarithmic negativity}

Logarithmic negativity \cite{neg4, neg5} is an entanglement measure that originates from the partial transposition criterion \cite{peres1996}. It is a necessary and sufficient condition for quantifying entanglement for arbitrary two-qubit states. For any two-qubit state $\rho_{AB}$, logarithmic negativity $\mathcal{E}$ can be defined as
\begin{equation*}
    \mathcal{E}(\rho_{AB}) = \log_{2}[2N(\rho_{AB}) + 1],
\end{equation*}
where $N$ is the negativity defined as 
\begin{equation*}
    N(\rho_{AB}) = \frac{||\rho_{AB}^{T_{A}}||_{1} - 1}{2}.
\end{equation*}
Here, $||\rho||_{1}$ is the trace-norm of the matrix $\rho$ defined as, $||\rho||_{1} = \mathbf{tr} \sqrt{\rho^{\dagger}\rho}$ and $T_{A}$ is the partial transpose of $\rho_{AB}$ with respect to $A$.

\section{Concurrence} 
\label{Appendis:C}

Concurrence \cite{wootters2001} quantifies the amount of entanglement present in an arbitrary two-qubit state. Given a two-qubit density matrix, $\rho_{AB}$, the concurrence is defined as 
\begin{equation} 
 C(\rho_{AB})=\max \{0, \lambda_1 - \lambda_2 - \lambda_3 - \lambda_4\},
\end{equation}
where the $\lambda_i$ are the eigenvalues of the Hermitian  matrix, with $ R=\sqrt{\sqrt{\rho}~\widetilde{\rho}~\sqrt{\rho}}$ satisfying the order $\lambda_1\geq \lambda_2\geq \lambda_3\geq \lambda_4$. Here $ \widetilde{\rho}=(\sigma_y \otimes \sigma_y) \rho^\ast (\sigma_y \otimes \sigma_y)$ with  $\rho^*$  the complex conjugate of $\rho$ in the computational basis.

\section{Monogamy Score}
\label{Appendis:C} 

The monogamy score quantifies the distribution of the entanglement among $N$-parties of a quantum state, $\rho_{1 2,...,N}$. Monogamy of entanglement states that if entanglement between two parties is maximum, they cannot share any amount of entanglement with other parties. To find the trade-off relations between entanglement content among parties, we use concurrence, $C_{ij}$ between spins $i$ and $j$ as a bipartite entanglement measure. By considering the first spin as the node and calculating the entanglement shared between the first spin and rest of the system, denoted by $C(\rho_{1:\text{rest}})$, we can define the monogamy score as 
\begin{equation}
    \delta_{C^{2}} = C^2(\rho_{1:\text{rest}})-\sum_{i=2}^N C^2(\rho_{1i}),
\end{equation}
where $C^2(\rho_{1i})$ denotes the concurrence between the first and any arbitrary site, $i$. Note that $C^2(\rho_{1,\text{rest}}) \leq \log_{2}d_{1}$, where $d_{1}$ is the dimension of the first spin which is unity for a two-qubit case. Similarly, $C^2(\rho_{1i}) \leq 1$. It was shown that for ani arbitrary $N$-party state \cite{coffman2000, ckw2}, $\delta_{C^2} \geq 0$.

\end{document}